\documentclass[preprint,showpacs,preprintnumbers,amsmath,amssymb]{revtex4}
\usepackage{graphicx}
\usepackage{dcolumn}
\usepackage{bm}

\font\scripti=cmmi7
\font\scriptscripti=cmmi5
\def\sib#1{\setbox0 = \hbox{\scripti #1}
  \kern-.02em\copy0\kern-\wd0
  \kern.04em\box0} 
\def\ssib#1{\setbox0 = \hbox{\scriptscripti #1}
  \kern-.02em\copy0\kern-\wd0
  \kern.04em\box0} 
\font\tenib=cmmib10 
\skewchar\tenib='177 \skewchar\tenib='177 \skewchar\tenib='177
\def\pbold#1{\setbox0 = \hbox{$ #1 $}
  \kern-.022em\copy0\kern-\wd0
  \kern.011em\copy0\kern-\wd0
  \kern.011em\copy0\kern-\wd0
  \kern.011em\copy0\kern-\wd0
  \kern.011em\box0} 

\usepackage{graphicx}
\usepackage{dcolumn}
\usepackage{bm}

\def\op{\Delta}

\def\epl{\varepsilon}
\def\up{\uparrow}
\def\dwn{\downarrow}

\def\lesssim{\ \raise.3ex\hbox{$<$}\kern-0.8em\lower.7ex\hbox{$\sim$}\ }
\def\gesim{\ \raise.3ex\hbox{$>$}\kern-0.8em\lower.7ex\hbox{$\sim$}\ }

\begin{document}
\title{Strong-coupling corrections to spin susceptibility in the BCS-BEC crossover regime of a superfluid Fermi gas}
\author{Hiroyuki Tajima, Ryo Hanai, and Yoji Ohashi}
\affiliation{Department of Physics, Keio University, 3-14-1 Hiyoshi, Kohoku-ku, Yokohama 223-8522, Japan}
\date{\today}
\begin{abstract}
We theoretically investigate the uniform spin susceptibility $\chi$ in the superfluid phase of an ultracold Fermi gas in the BCS (Bardeen-Cooper-Schrieffer)-BEC (Bose-Einstein condensation) crossover region. In our previous paper [H. Tajima, {\it et. al.}, Phys. Rev. A {\bf 89}, 033617 (2014)], including pairing fluctuations within an extended $T$-matrix approximation (ETMA), we showed that strong pairing fluctuations cause the so-called spin-gap phenomenon, where $\chi$ is anomalously suppressed even in the normal state near the superfluid phase transition temperature $T_{\rm c}$. In this paper, we extend this work to the superfluid phase below $T_{\rm c}$, to clarify how this many-body phenomenon is affected by the superfluid order. From the comparison of the ETMA $\chi$ with the Yosida function describing the spin susceptibility in a weak-coupling BCS superfluid, we identify the region where pairing fluctuations crucially affect this magnetic quantity below $T_{\rm c}$ in the phase diagram with respect to the strength of a pairing interaction and the temperature. This spin-gap regime is found to be consistent with the previous pseudogap regime determined from the pseudogapped density of states. We also compare our results with a recent experiment on a $^6$Li Fermi gas. Since the spin susceptibility is sensitive to the formation of spin-singlet preformed pairs, our results would be useful for the study of pseudogap physics in an ultracold Fermi gas on the viewpoint of the spin degrees of freedom.
\end{abstract}
\maketitle
\par
\section{Introduction}
\par
Since the achievement of the BCS (Bardeen-Cooper-Schrieffer)-BEC (Bose-Einstein condensation) crossover phenomenon\cite{Eagles,Leggett,NSR,SadeMelo,Perali2,Ohashi,Ohashi2003} in ultracold $^{40}$K\cite{Jin} and $^6$Li\cite{Zwierlein,Kinast,Jochim} Fermi gases, physical properties of this strongly interacting Fermi system have attracted much attention\cite{Bloch,Gurarie,Haussmann2,Giorgini,Chin}, especially in the unitary regime\cite{Heiselberg,Carlson,Ho,Hu2,Nishida,Luo,Horikoshi,Ku,Pantel,Palestini,Mink}. In addition, the photoemission-type experiments on $^{40}$K Fermi gases\cite{Stewart,Gaebler,Perali,Kohl,Sagi} have stimulated the research for the pseudogap phenomenon in the BCS-BEC crossover region\cite{Tsuchiya1,Tsuchiya2,Tsuchiya3,Watanabe1,Watanabe2,Watanabe3,Mueller,Chen,Hu,Magierski,Su,Bulgac,Boettcher}. 
\par
Although the pseudogap has extensively been discussed in the under-doped regime of high-$T_{\rm c}$ cuprates\cite{Renner,Krasnov,Fauque,Fischer,Ma,Hashimoto,Hashimoto2}, the complexity of this strongly correlated electron system still prevents us from the complete understanding of this many-body phenomenon\cite{Pines,Kamp,Chakravarty,Yanase,Shen,Kivelson,Lee2}. In contrast, an ultracold Fermi gas in the BCS-BEC crossover region is simply dominated by pairing fluctuations. Thus, once the pseudogap is observed in this atomic system, one can immediately conclude that it originates from preformed Cooper pairs. This so-called preformed pair scenario\cite{Yanase} is also a candidate for the pseudogap mechanism in high-$T_{\rm c}$ cuprates, so that this observation would also contribute to the assessment of this scenario. At present, it has theoretically been pointed out\cite{Perali,Tsuchiya3,Chen,Hu,Magierski} that the deviation of the photoemission spectrum from the free particle dispersion observed in a $^{40}$K unitary Fermi gas\cite{Stewart,Gaebler} may be an indirect evidence for the pseudogap phenomenon. However, since the current photoemission-type experiment does not have energy resolution enough to construct the single-particle density of states $\rho(\omega)$, a dip structure in $\rho(\omega)$ (which is the most direct evidence of the pseudogap phenomenon) has not been confirmed yet. Because of this, it is still in debate whether pairing fluctuations in an ultracold Fermi gas really cause the pseudogap phenomenon or not\cite{Nascimbene2,Nascimbene,Navon}.
\par
In a previous paper\cite{Tajima}, including strong pairing fluctuations within an extended $T$-matrix approximation (ETMA)\cite{Kashimura1,Kashimura2,Hanai}, we examined strong-coupling corrections to the spin susceptibility $\chi$, which is experimentally accessible\cite{Sanner,Sommer,Esslinger,Lee}, in the normal state near the superfluid phase transition temperature $T_{\rm c}$, to clarify that this magnetic quantity is useful for the study of pseudogap physics in an ultracold Fermi gas with an $s$-wave pairing interaction. (Here, spin $\sigma=\uparrow,\downarrow$ is actually pseudospin, describing two atomic hyperfine states contributing to the pair formation.) The formation of spin-singlet preformed Cooper pairs was shown to suppress $\chi$ below the so-called spin-gap temperature $T_{\rm s}$. In the BCS-BEC crossover region, this characteristic temperature $T_{\rm s}$ was found to be comparable to the pseudogap temperature $T^*$, which is determined as the temperature below which a dip structure appears in $\rho(\omega)$ around $\omega=0$. We also showed that the calculated spin-susceptibility agrees well with the recent experiment on a $^6$Li Fermi gas\cite{Sanner}, indicating that the observed small $\chi$ may be due to the formation of spin-singlet preformed Cooper pairs near $T_{\rm c}$. We briefly note that, although the spin-gap phenomenon has also been discussed in high-$T_{\rm c}$ cuprates\cite{Yoshinari}, the origin of this anomaly, as well as relation to the pseudogap phenomenon, are still controversial in this electron system. In contrast, the pseudogap and the spin-gap in an ultracold Fermi gas are different aspects of the same fluctuation phenomenon, where preformed Cooper pairs play crucial roles.
\par
While the pseudogap is usually discussed in the normal state, it is an interesting problem whether or not this phenomenon also occurs in the superfluid phase below $T_{\rm c}$. This problem was recently examined on the viewpoint of the superfluid density of states $\rho(\omega)$\cite{Watanabe1,Watanabe2}, and it was clarified that the pseudogap in $\rho(\omega)$ remains just below $T_{\rm c}$, to continuously change to the ordinary BCS-type superfluid gap at low temperatures, reflecting the suppression of pairing fluctuations by the superfluid order. However, such ``pseudogapped" superfluid density of states expected near $T_{\rm c}$ is difficult to observe in the current stage of cold Fermi gas physics. On the other hand, since the spin susceptibility $\chi$ is experimentally accessible below $T_{\rm c}$\cite{Sanner}, the confirmation of this predicted superfluid pseudogap phenomenon is promising through the spin-gap phenomenon appearing in $\chi$. 
\par
Motivated by this expectation, in this paper, we investigate the uniform spin susceptibility $\chi$ in the BCS-BEC crossover regime of a superfluid Fermi gas. Extending our previous work for the normal state\cite{Tajima} to the superfluid phase below $T_{\rm c}$, we show that effects of pairing fluctuations on $\chi$ appears as deviation from the Yosida function $\chi_{\rm Yosida}$\cite{Yosida}, describing the spin susceptibility in the ordinary weak-coupling BCS superfluid. Using this deviation, we determine the region where pairing fluctuations crucially affect $\chi$ below $T_{\rm c}$, in the phase diagram with respect to the strength of a pairing interaction and the temperature. This region is found to be consistent with the ``pseudogapped superfluid regime" that was previously predicted from an analysis on the superfluid density of states\cite{Watanabe1,Watanabe2}.
\par
This paper is organized as follows. In Sec. II, we explain an extended $T$-matrix approximation (ETMA), to evaluate the spin susceptibility $\chi$ in a superfluid Fermi gas. In Sec. III, we show the calculated $\chi$ as a function of temperature in the whole BCS-BEC crossover region. Here, we also compare our results with the recent experiment on a $^6$Li Fermi gas\cite{Sanner}. In Sec. IV, we compare $\chi$ with the Yosida function, to evaluate effects of pairing fluctuations from their difference. Using this, we identify the superfluid spin-gap regime, where pairing fluctuations crucially affect $\chi$ even in the superfluid phase, in the phase diagram with respect to the interaction strength and the temperature. We also discuss how this region is related to the previous pseudogap regime which was determined from the pseudogapped superfluid density of states\cite{Watanabe1,Watanabe2}. In this paper, we take $\hbar=k_{\rm B}=1$, and the system volume $V$ is taken to be unity, for simplicity.
\par
\section{Formulation}
\par
We consider a three dimensional uniform superfluid Fermi gas, described by the ordinary BCS model. Under the Nambu representation\cite{Watanabe1,Schrieffer,Ohashi2,Pieri,Fukushima}, the BCS Hamiltonian is written as
\begin{eqnarray}
H=\sum_{\bm p}\Psi^{\dag}_{\bm p}\left[\xi_{\bm p}\tau_{3}-\op\tau_{1}-h\right]\Psi_{\bm p} -U\sum_{\bm q}\rho_{+}(\bm{q})\rho_{-}(-\bm{q}).
\label{eq2-1}
\end{eqnarray}
Here, 
\begin{eqnarray}
\Psi_{\bm p}=
\left(
\begin{array}{c}
c_{\bm{p},\up} \\
c^{\dag}_{-\bm{p},\dwn}
\end{array}
\right)
\label{eq2-1b}
\end{eqnarray}
is the two-component Nambu field, where $c_{\bm{p},\sigma}$ is the annihilation operator of a Fermi atom with pseudospin $\sigma=\uparrow,\downarrow$, describing two atomic hyperfine states. The Pauli matrices $\tau_j$ ($j=1,2,3$) act on particle-hole space. $\xi_{\bm p}=\varepsilon_{\bm p}-\mu=p^2/(2m)-\mu$ is the kinetic energy of a Fermi atom, measured from the Fermi chemical potential $\mu$, where $m$ is an atomic mass. Although we consider an unpolarized Fermi gas, we have added an infinitesimal effective magnetic field $h$ to the model Hamiltonian in Eq. (\ref{eq2-1}), to calculate the spin susceptibility $\chi$ (see Eq. (\ref{eq2-9})). 
\par
In Eq. (\ref{eq2-1}), the ordinary contract-type $s$-wave pairing interaction,
\begin{equation}
H_{\rm BCS}=-U
\sum_{{\bm p},{\bm p}',{\bm q}}
c_{{\bm p}+{\bm q}/2,\uparrow}^\dagger 
c_{-{\bm p}+{\bm q}/2,\downarrow}^\dagger 
c_{-{\bm p}'+{\bm q}/2,\downarrow} 
c_{-{\bm p}+{\bm q}/2,\uparrow}
\label{eq2-1c}
\end{equation}
(where $-U$ ($<0$) is the interaction strength), has been divided into the mean-field term $\Delta\tau_1$ with the superfluid order parameter $\Delta$, and the last correction term to the mean-field approximation. Here, 
\begin{equation}
\rho_\pm ({\bm q})=
\frac{1}{2}[\rho_1({\bm q})\pm i\rho_2({\bm q})]
=
\sum_{\bm p}
\Psi_{\bm{p}+\bm{q}/2}^{\dag}\tau_{\pm}\Psi_{\bm{p}-\bm{q}/2}
\label{eq2-1d}
\end{equation}
is the generalized density operator\cite{Watanabe1,Ohashi2,Fukushima}, describing fluctuations of the superfluid order parameter $\Delta$, where $\tau_\pm=[\tau_1\pm i\tau_2]/2$. Since $\Delta$ is chosen to be parallel to the $\tau_1$ component in Eq. (\ref{eq2-1}), $\rho_1({\bm q})=\sum_{\bm p}\Psi_{\bm{p}+\bm{q}/2}^{\dag}\tau_1\Psi_{\bm{p}-\bm{q}/2}$ and $\rho_2({\bm q})=\sum_{\bm p}\Psi_{\bm{p}+\bm{q}/2}^{\dag}\tau_2\Psi_{\bm{p}-\bm{q}/2}$ in Eq. (\ref{eq2-1d}) physically mean amplitude fluctuations and phase fluctuations of the order parameter, respectively\cite{note1}. Substituting the first expression in Eq. (\ref{eq2-1d}) into the last term in Eq. (\ref{eq2-1}), we find that this term is written as the sum of the interaction between amplitude fluctuations, and that between phase fluctuations of the superfluid order parameter.
\par
Since the BCS Hamiltonian in Eq. (\ref{eq2-1}) involves the ultraviolet divergence, we need to eliminate this singularity. As usual, this can be achieved by measuring the interaction strength in terms of the $s$-wave scattering length $a_s$, which is related to $-U$ as,
\begin{equation}
{4\pi a_s \over m}=
-{U \over \displaystyle 1-U\sum_{\bm p}^{p_{\rm c}}{1 \over 2\varepsilon_{\bm p}}},
\label{eq2-2}
\end{equation}
where $p_{\rm c}$ is a cutoff momentum. In this scale, the weak coupling BCS regime and the strong-coupling BEC regime are described as $(k_{\rm F}a_{s})\lesssim -1$, and $(k_{\rm F}a_{s})^{-1}\gesim 1$, respectively (where $k_{\rm F}$ is the Fermi momentum). The region, $-1\lesssim (k_{\rm F}a_{s})^{-1}\lesssim 1$, is referred to as the BCS-BEC crossover region.
\par
\begin{figure}[t]
\begin{center}
\includegraphics[width=8cm]{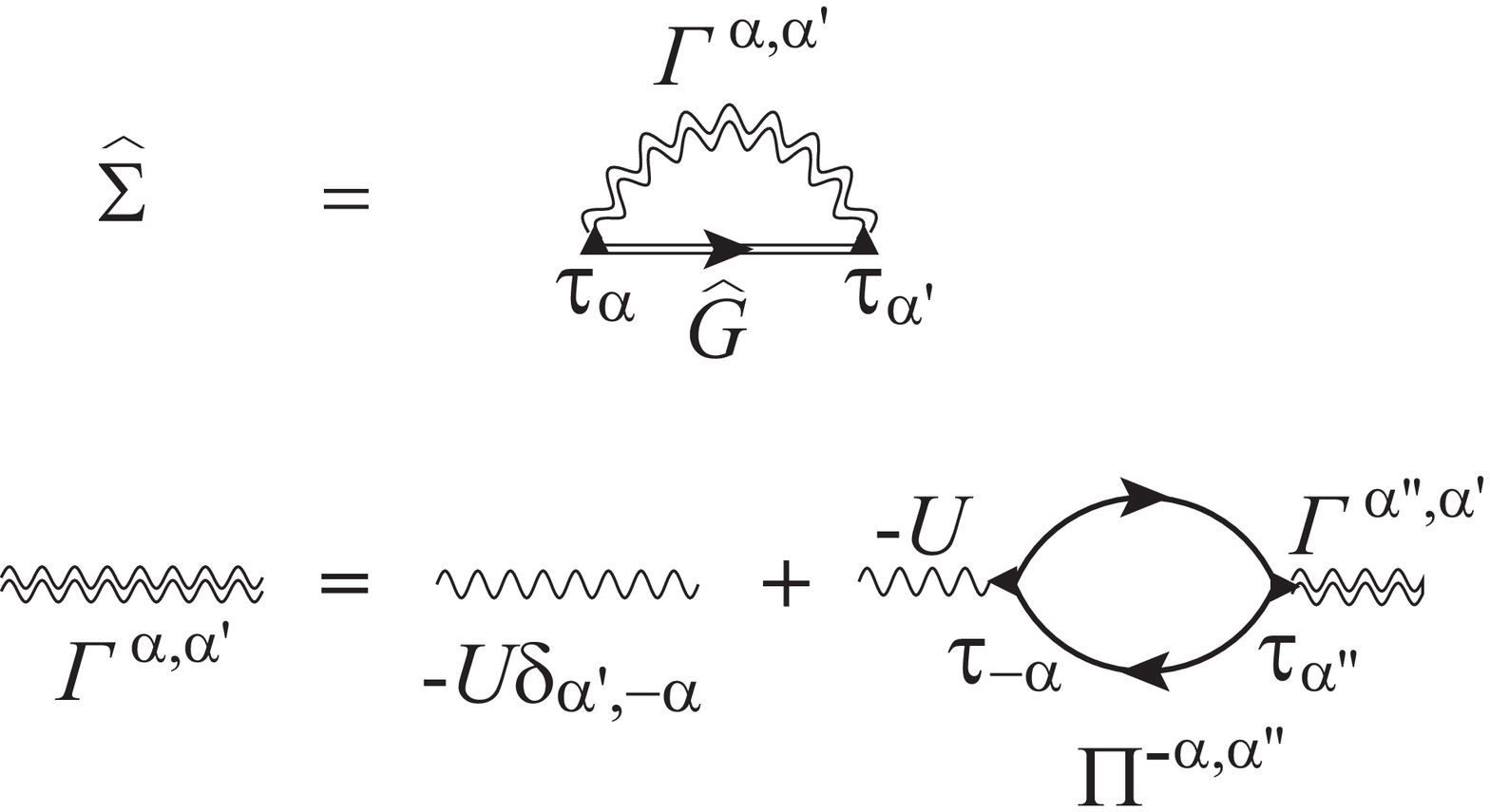}
\end{center}
\caption{Self-energy correction ${\hat \Sigma}({\bm p},i\omega_n)$ in ETMA. The double solid line is the ETMA Green's function ${\hat G}$ in Eq. (\ref{eq2-3}). The single solid line is the mean-field BCS Green's function ${\hat G}_0$ in Eq. (\ref{eq2-6b}). The double wavy line represents the ETMA particle-particle scattering matrix ${\hat \Gamma}({\bm q},i\nu_n)$ in Eq. (\ref{eq2-5}). The single wary line is the pairing interaction $-U$.}
\label{fig1}
\end{figure}
\par
Single-particle properties in the superfluid phase is conveniently described by the $2\times 2$ matrix single-particle thermal Green's function, given by
\begin{equation}
\hat{G}({\bm p},i\omega_n)=
{1 \over
i\omega_n-\xi_{\bm p}\tau_3+\Delta\tau_1+h-\hat{\Sigma}({\bm p},i\omega_n)},
\label{eq2-3}
\end{equation}
where $\omega_n$ is the fermion Matsubara frequency, and the $2\times 2$ matrix self-energy $\hat{\Sigma}({\bm p},i\omega_n)$ describes strong-coupling corrections. In this paper, we deal with $\hat{\Sigma}({\bm p},i\omega_n)$ within an extended $T$-matrix approximation (ETMA)\cite{Kashimura1,Tajima,Kashimura2,Hanai}, which is diagrammatically described as Fig. \ref{fig1}. Summing up these diagrams, we have
\begin{equation}
\hat{\Sigma}({\bm p},i\omega_n)
=-T\sum_{{\bm q},\nu_n}\sum_{\alpha,\alpha'=\pm}
{\hat \Gamma}^{\alpha,\alpha'}({\bm q},i\nu_n)
\tau_\alpha\hat{G}({\bm p}+{\bm q},i\omega_n+i\nu_n)\tau_{\alpha'},
\label{eq2-4}
\end{equation}
where $\nu_n$ is the boson Matsubara frequency, and
\begin{eqnarray}
\hat{\Gamma}({\bm q},i\nu_n)&=&
\left(\begin{array}{cc}
\Gamma^{-+}({\bm q},i\nu_n) & \Gamma^{--}({\bm q},i\nu_n) \\
\Gamma^{++}({\bm q},i\nu_n) & \Gamma^{+-}({\bm q},i\nu_n) 
\end{array}\right) \cr
&=&-U\left[1+U
\left
(\begin{array}{cc}
\Pi^{-+}({\bm q},i\nu_n) & \Pi^{--}({\bm q},i\nu_n) \\
\Pi^{++}({\bm q},i\nu_n) & \Pi^{+-}({\bm q},i\nu_n) 
\end{array}
\right)\right]^{-1}
\label{eq2-5}
\end{eqnarray}
is the particle-particle scattering matrix. In Eq. (\ref{eq2-5}),
\begin{equation}
\Pi^{\alpha,\alpha'}({\bm q},i\nu_n)=
T\sum_{{\bm p},i\omega_n}
{\rm Tr}\left[\tau_\alpha\hat{G}^0({\bm p}+{\bm q},i\omega_n+i\nu_n)
\tau_{\alpha'}G^0({\bm p},i\omega_n)\right]
\label{eq2-6}
\end{equation}
is the pair-correlation function, where
\begin{equation}
\hat{G}_0({\bm p},i\omega_n)=
{1 \over i\omega_n-\xi_{\bm p}\tau_3+\Delta\tau_1+h}
\label{eq2-6b}
\end{equation}
is the single-particle thermal Green's function in the mean-field BCS theory\cite{Schrieffer}. Since we are choosing the superfluid order parameter $\Delta$ to be parallel to the $\tau_1$ component in Eq. (\ref{eq2-1}), the pair correlation function in Eq. (\ref{eq2-6}) with $\alpha=\alpha'=1$ physically describes amplitude fluctuations of the superfluid order parameter\cite{Ohashi2,Fukushima}. $\Pi^{22}$ and $\Pi^{12}$ ($=-\Pi^{21}$) describe phase fluctuations of the superfluid order parameter, and coupling between phase and amplitude fluctuations, respectively. Thus, $\Pi^{\pm,\pm}$ in the particle-particle scattering matrix in Eq. (\ref{eq2-5}) involves these fluctuation effects existing in the superfluid phase below $T_{\rm c}$. 
\par
\begin{figure}[t]
\begin{center}
\includegraphics[width=8cm]{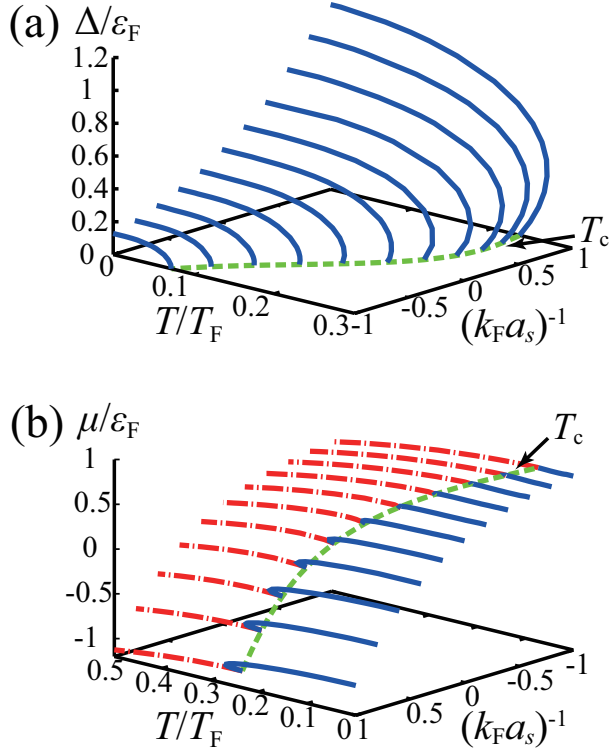}
\end{center}
\caption{(color online) Self-consistent solutions for the coupled gap equations (\ref{eq2-7}) with the number equation $N$. (a) Superfluid order parameter $\Delta$. (b) Fermi chemical potential $\mu$. In these figures, the first-order behavior seen in the crossover region is an artifact of ETMA\cite{Watanabe1,Fukushima}.}
\label{fig2}
\end{figure}
\par
The gap equation for $\Delta$ is obtained from the condition that the ETMA particle-particle scattering matrix ${\hat \Gamma}({\bm q},i\nu_n)$ in Eq. (\ref{eq2-5}) has a pole at ${\bm q}=\nu_n=0$ (which guarantees the required gapless Goldstone mode). The resulting equation has the same form as the ordinary BCS gap equation, as
\begin{equation}
1=-\frac{4\pi a_{s}}{m}\sum_{\bm{p}}
\left[
{1 \over 2E_{\bm p}}\tanh{E_{\bm p} \over 2T}
-\frac{1}{2\epl_{\bm{p}}}
\right],
\label{eq2-7}
\end{equation}
where $E_{\bm p }=\sqrt{\xi_{\bm p}^2+\Delta^2}$ is the Bogoliubov single-particle excitation spectrum. We numerically solve the gap equation (\ref{eq2-7}), together with the equation for the total number $N=N_\uparrow+N_\downarrow$ of Fermi atoms, to self-consistently determine $\Delta$ and $\mu$. In this number equation, $N_\sigma$ is the number of Fermi atoms with pseudospin $\sigma$, given by
\begin{eqnarray}
\begin{array}{l}
\displaystyle
N_\uparrow = T\sum_{{\bm p},\omega_n}G_{11}({\bm p},i\omega_n),
\\
\displaystyle
N_\downarrow = 
\sum_{\bm p}1-T\sum_{{\bm p},\omega_n}G_{22}({\bm p},i\omega_n),
\end{array}
\label{eq2-8}
\end{eqnarray}
where $G_{ii}$ is the diagonal ($i,i$)-component of the ETMA Green's function in Eq. (\ref{eq2-3}). 
\par
Figure \ref{fig2} shows the self-consistent solutions for the superfluid order parameter $\Delta$, as well as the Fermi chemical potential $\mu$. These results will be used in evaluating the spin susceptibility $\chi$. 
\par
One sees in Fig. \ref{fig2}(a) that $\Delta$ exhibits the first-order behavior just below $T_{\rm c}$, when the pairing interaction becomes strong. This is, however, an artifact of ETMA. The same problem has also been known in other diagrammatic strong-coupling theories\cite{Watanabe1,Fukushima,Haussmann}, where the origin of this deficiency is considered to be incomplete treatment of an effective interaction between Cooper pairs\cite{OhashiJPSJ}. Although it is a crucial issue to overcome this problem, we leave this as a future problem, and we examine strong-coupling corrections to $\chi$ below $T_{\rm c}$ within the framework of ETMA in this paper.
\par
\begin{figure}[t]
\begin{center}
\includegraphics[width=10cm]{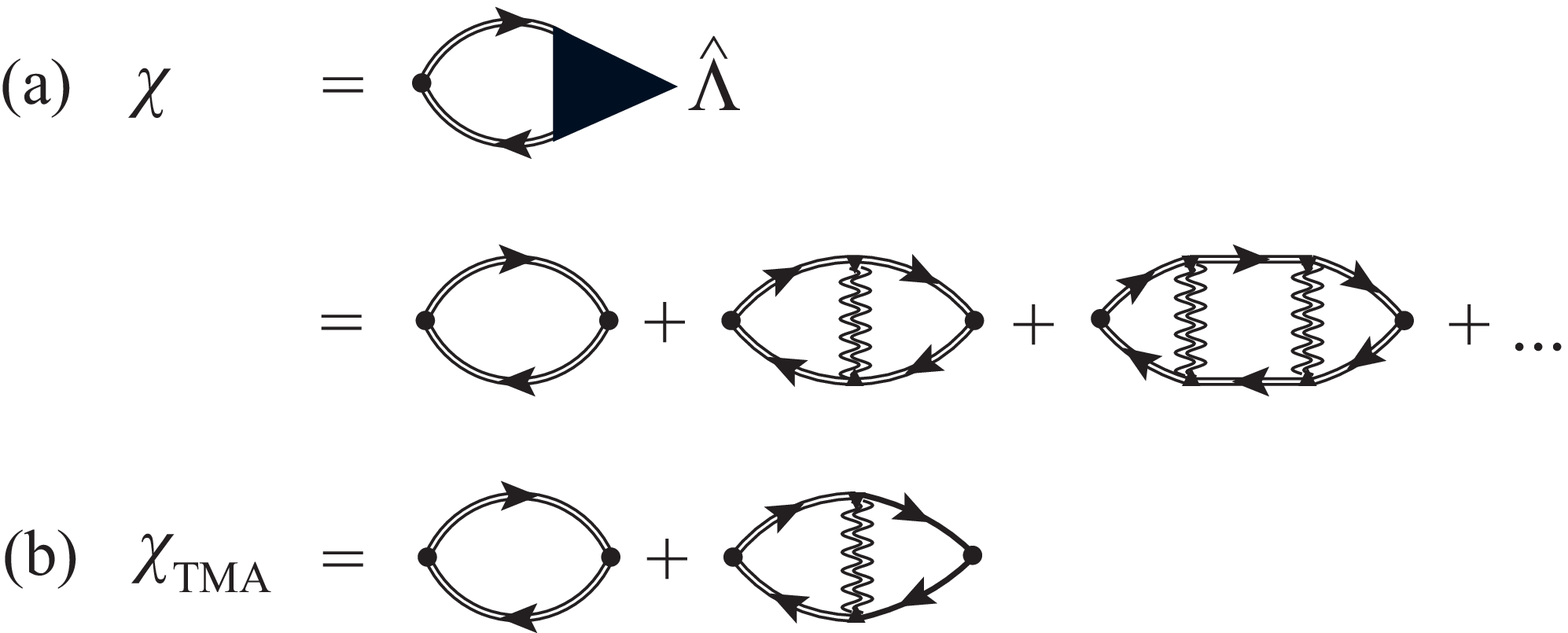}
\end{center}
\caption{(a) Diagrammatic representation of ETMA spin susceptibility $\chi$. The filled circle and filled triangle ${\hat \Lambda}$ are the bare and the dressed spin-vertex, respectively. The double solid line is the dressed Green's function ${\hat G}$. In this figure, the second line shows vertex corrections to $\chi$ in ETMA, where the double wavy line is the particle-particle scattering matrix ${\hat \Gamma}$. (b) Spin susceptibility in the ordinary (non-self-consistent) $T$-matrix approximation (TMA). The single solid line is the mean-field BCS Green's function ${\hat G}_0$.}
\label{fig3}
\end{figure}
\par
The spin susceptibility $\chi$ is given by
\begin{equation}
\chi=\left. \frac{\partial (N_{\up}-N_{\dwn})}{\partial h}\right|_{h\rightarrow 0}=
\lim_{h\to 0}{N_\uparrow-N_\downarrow \over h}.
\label{eq2-9}
\end{equation}
Although $\chi$ can also be calculated by using the diagrammatic technique shown in Fig. \ref{fig3}(a), the advantage of using Eq. (\ref{eq2-9}) is that one can automatically treat the self-energy correction ${\hat \Sigma}$ and the spin-vertex correction in a consistent manner. Indeed, substituting Eqs. (\ref{eq2-8}) into the first expression in (\ref{eq2-9}), we obtain
\begin{eqnarray}
\chi=-T\sum_{{\bm p},\omega_n}{\rm Tr}
\left[
\hat{G}^{2}({\bm p},i\omega){\hat \Lambda}({\bm p},i\omega_n)
\right],
\label{eq2-9b}
\end{eqnarray}
where the dressed spin vertex ${\hat \Lambda}({\bm p},i\omega)$ is related to the self-energy ${\hat \Sigma}({\bm p},i\omega_n)$ as
\begin{equation}
{\hat \Lambda}({\bm p},i\omega_n)=
1-
\left.
{\partial {\hat \Sigma}({\bm p},i\omega_n) \over \partial h}
\right|_{h\to 0}.
\label{eq2-9c}
\end{equation}
Equation (\ref{eq2-9c}) is just the Ward identity for the spin-vertex correction ${\hat \Lambda}({\bm p},i\omega)$\cite{Ward}, that must be satisfied in any consistent theory. 
\par
As pointed out in Ref. \cite{Kashimura1}, ETMA can overcome the serious problem that $\chi$ unphysically becomes negative in the unitary regime in the the ordinary (non-self-consistent) $T$-matrix approximation (TMA)\cite{Kashimura1}, as well as in the strong-coupling theory developed by Nozi\`eres and Schmitt-Rink\cite{Liu,Parish}. To see this in a simple manner, we note that, in ETMA, Eq. (\ref{eq2-9c}) gives the infinite series of bubble diagram shown in the second line in Fig. \ref{fig3}(a)\cite{note2}. For simplicity, approximating the particle-particle scattering matrix to the bare interaction as ${\hat \Gamma}=-U{\hat 1}$ (where ${\hat 1}$ is the $2\times 2$ unit matrix), one obtains the RPA (random phase approximation) type expression for the ETMA spin susceptibility as,
\begin{equation}
\chi\simeq 
{\chi_{\rm DOS}
\over
\displaystyle
1+{U \over 2}\chi_{\rm DOS}}.
\label{eq2-9d}
\end{equation}
Since the so-called DOS (density of states) diagram $\chi_{\rm DOS}$, which is given by the first term in the second line in Fig. \ref{fig3}(a), is always positive, the positivity of $\chi$ is guaranteed. In contrast, in the case of TMA, where the self-energy is given by replacing the dressed Green's function ${\hat G}$ appearing in Eq. (\ref{eq2-4}) with the mean-field BCS one ${\hat G}_0$ in Eq. (\ref{eq2-6b}), the RPA series is found to be truncated to the first order, as shown in Fig. \ref{fig3}(b). Employing the same treatment (${\hat \Gamma}=-U{\hat 1}$) as in the ETMA case, one obtains the TMA spin susceptibility $\chi_{\rm TMA}$ as,
\begin{equation}
\chi_{\rm TMA}\simeq
\chi_{\rm DOS}
\left[
1-{U \over 2}\chi_{\rm DOS}^0
\right],
\label{eq2-9e}
\end{equation}
where $\chi_{\rm DOS}^0$ is the DOS diagram where the mean-field BCS Green's function ${\hat G}_0$ is used for ${\hat G}$. $\chi_{\rm TMA}$ in Eq. (\ref{eq2-9e}) unphysically becomes negative, when $(U/2)\chi_{\rm DOS}^0>1$. 
\par
In this paper, we numerically evaluate Eq. (\ref{eq2-9}), by taking a small but finite value of the effective magnetic field, $h/\varepsilon_{\rm F}=O(10^{-2}\sim 10^{-3})$, where $\varepsilon_{\rm F}$ is the Fermi energy. We have confirmed the linear-$h$ dependence of $N_\uparrow-N_\downarrow$ in this regime.
\par
\section{Spin susceptibility in the BCS-BEC crossover regime of a superfluid Fermi gas}
\par
\begin{figure}[t]
\begin{center}
\includegraphics[width=8cm]{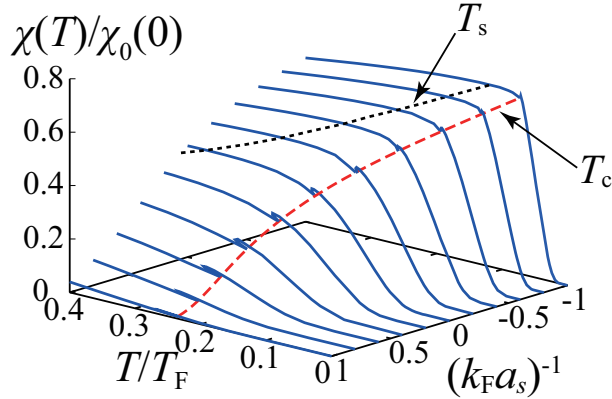}
\end{center}
\caption{(color online) Calculated spin susceptibility $\chi$ in the BCS-BEC crossover regime of an ultracold Fermi gas, normalized by the value $\chi_0(0)$ in an ideal Fermi gas at $T=0$. The same normalization is also used in Figs. \ref{fig5} and \ref{fig6}. The dashed line is $T_{\rm c}$, and the dotted line is the spin-gap temperature $T_{\rm s}$, which is determined as the temperature at which $\chi$ takes a maximum value. $T_{\rm F}$ is the Fermi temperature. The singular behavior seen around $T_{\rm c}$ is an artifact of ETMA.}
\label{fig4}
\end{figure}
\begin{figure}[t]
\begin{center}
\includegraphics[width=6.15cm]{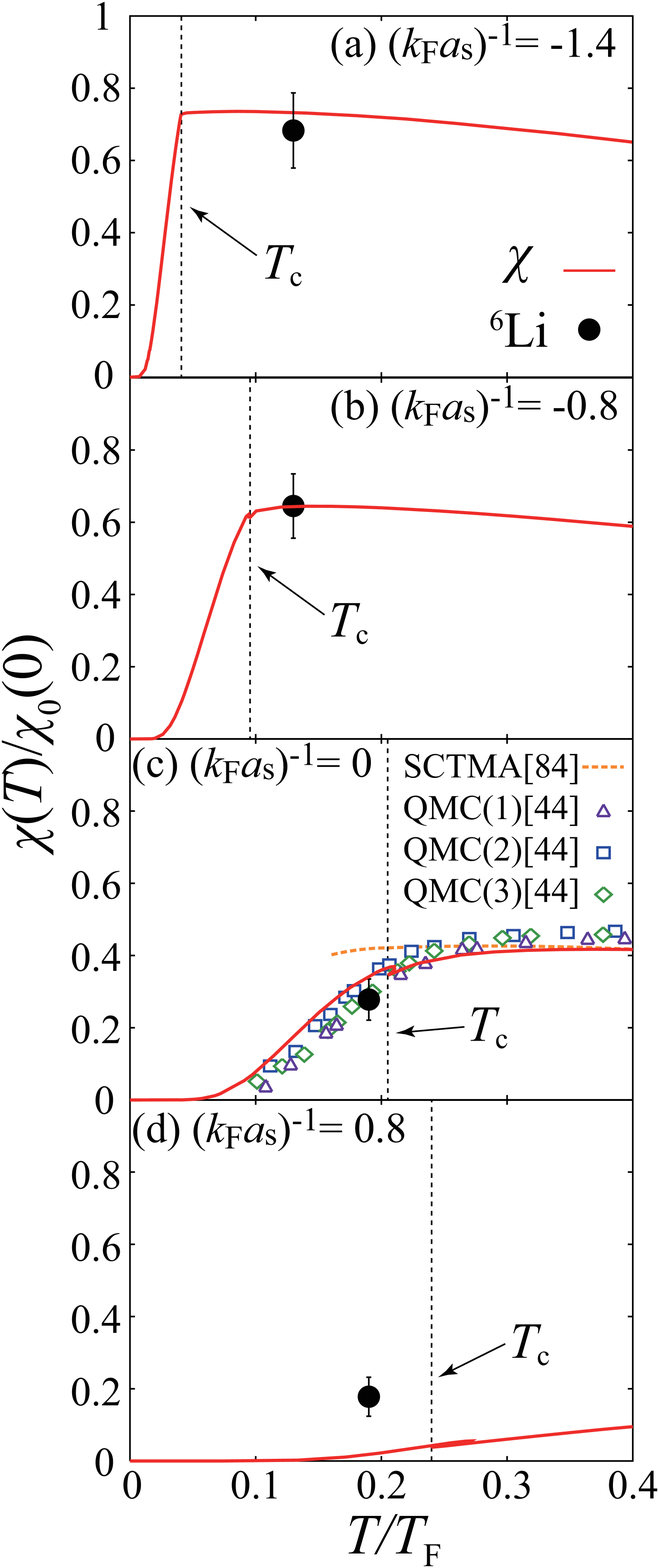}
\end{center}
\caption{(color online). Calculated spin susceptibility $\chi$ as a function of temperature. The filled circles are experimental data on a $^6$Li Fermi gas\cite{Sanner}. The dotted line shows $T_{\rm c}$ at each interaction strength. In panel (c), we also compare our result with the previous work by the self-consistent $T$-matrix approximation (SCTMA)\cite{Enss} as well as quantum Monte-Carlo simulations\cite{Bulgac}, where $(N_x,n)=(8,0.08)$ (QMC(1)), $(10,0.04)$ (QMC(2)), and $(12,0.03)$ (QMC(3)) (where $N_x$ and $n$ are the lattice size and the filling density per lattice site, respectively).}
\label{fig5}
\end{figure}

Figure \ref{fig4} shows the uniform spin susceptibility $\chi$ in the BCS-BEC crossover regime of an ultracold Fermi gas. For clarity, we also summarize in Fig. \ref{fig5} the detailed temperature dependence of $\chi$ at some interaction strengths. In these figures, the singularity around $T_{\rm c}$ is an artifact of ETMA, as mentioned previously. Apart from this, in the whole BCS-BEC crossover region, $\chi$ is found to decrease with decreasing the temperature below the spin-gap temperature $T_{\rm s}$ (which is determined as the temperature at which $\chi$ takes a maximum value\cite{Tajima}). In the BCS side ($(k_{\rm F}a_s)^{-1}<0$), this decrease is more remarkable in the superfluid phase below $T_{\rm c}$ than in the spin-gap regime, $T_{\rm c}\le T\le T_{\rm s}$. On the other hand, the temperature dependence of $\chi$ in the BEC side is not so sensitive to the superfluid instability as the weak-coupling case (apart from the unphysical singularity around $T_{\rm c}$), as shown in Fig. \ref{fig5}(d). 
\par
\par
\par
This insensitivity is deeply related to the fact that the superfluid instability in this regime is dominated by BEC of tightly bound molecules that have already been formed above $T_{\rm c}$\cite{Eagles,Leggett,NSR,SadeMelo}. Since the small but finite $\chi$ in Fig. \ref{fig5}(d) comes from partial dissociation of these molecules by thermal effects, the temperature dependence of $\chi$ is essentially unrelated to whether the system is in the normal state or in the superfluid phase in the strong-coupling BEC regime. 
\par
In the BCS side ($(k_{\rm F}a_s)^{-1}<0$), on the other hand, preformed Cooper pairs in the spin-gap regime ($T_{\rm c}\le T\le T_{\rm s}$) are actually fluctuating, in the sense that they frequently repeat pair-breaking and pair-formation. As a result, the suppression of spin degrees of freedom, as well as $\chi$, by these fluctuating spin-singlet preformed pairs is actually not so remarkable. However, once the system is in the superfluid phase, they start to become stable spin-singlet Cooper pairs, which enhances the suppression of $\chi$, as seen in Figs. \ref{fig5}(a) and (b).
\par
Figure \ref{fig5} also shows the recent experiment on a $^6$Li Fermi gas\cite{Sanner}. In addition to the weak-coupling case above $T_{\rm c}$ (Figs. \ref{fig5}(a) and (b)), the ETMA $\chi$ is also consistent with the experiment on a unitary Fermi gas below $T_{\rm c}$, as shown in Fig. \ref{fig5}(c). We briefly note that our result at the unitarity also agrees with the recent result in the self-consistent $T$-matrix approximation\cite{Enss} above $T_{\rm c}$, as well as the quantum Monte-Carlo simulation\cite{Bulgac}, as shown in Fig. \ref{fig5}(c). 
\par
However, Fig. \ref{fig5}(d) clearly shows that our result is much smaller than the experimental result at $T=0.19T_{\rm F}$, when $(k_{\rm F}a_s)^{-1}=0.8$ (where $T_{\rm F}$ is the Fermi temperature). In order to reproduce the observed large value of $\chi$ within ETMA, we need to take $T\simeq 0.9T_{\rm F}$, which is much higher than the superfluid phase transition temperature $T_{\rm c}\simeq 0.2T_{\rm F}$ expected in this regime. However, Ref. \cite{Sanner} reports that this measurement was done in the superfluid state with a finite condensate fraction.
\par
This discrepancy implies that ETMA underestimates the dissociation of Cooper pairs in the BEC regime. To examine this, when we deal with an ultracold Fermi gas in this regime as a simple Bose-Fermi mixture with $N_{\rm F}$ free Fermi atoms and $N_{\rm M}$ ideal spinless Bose molecules, the ETMA Fermi chemical potential $\mu(T=0.19T_{\rm F},(k_{\rm F}a_s)^{-1}=0.8 )=-0.53\varepsilon_{\rm F}$ gives the molecular dissociation rate $\alpha\equiv N_{\rm F}/N$ as,
\begin{equation}
\alpha={2\sum_{\bm p}f(\varepsilon_{\bm p}-\mu) \over N}=0.67\%,
\label{eqBF1}
\end{equation}
where $f(x)$ is the Fermi distribution function. This is relatively close to the value $\alpha=0.37\%$ which is obtained when one uses the expression $\mu=-1/(2ma_s^2)$ in the BEC limit. On the other hand, when we determine the value of $\mu$ so that the spin susceptibility in this model Bose-Fermi mixture\cite{Kubo,note3},
\begin{equation}
\chi_{\rm BF}=-2\sum_{\bm p}
{\partial f(\varepsilon_{\bm p}-\mu) \over \partial \varepsilon_{\bm p}},
\label{eqBF3}
\end{equation}
can reproduce the experimental result $\chi(T=0.19T_{\rm c})/\chi_0(0)=0.19$, one has $\alpha=7.5$\% (where $\chi_0(0)$ is the spin susceptibility in a free Fermi gas at $T=0$). This is much larger than the ETMA result in Eq. (\ref{eqBF1}), which implies that an additional strong depairing effect which is not taken into account in the present ETMA is necessary to explain this experiment\cite{Sanner}. Although clarifying the origin of this disrepancy is a crucial issue, we leave this as our future problem. In the next section, we examine strong-coupling corrections to $\chi$ in a superfluid Fermi gas within the present ETMA formalism.
\par
\par
\section{Phase diagram of an ultracold Fermi gas on the viewpoint of spin susceptibility}
\par
In the weak-coupling BCS theory, the spin susceptibility is known to be described by the Yosida function\cite{Yosida}, given by
\begin{eqnarray}
\chi_{\rm Yosida}=
-2\sum_{\bm p}
{\partial f(E_{\bm p}) \over \partial E_{\bm p}}
=
{1 \over 2T}\sum_{\bm p}{\rm sech}^2
\left({E_{\bm p} \over 2T}\right).
\label{eq3-1}
\end{eqnarray}
While the Fermi chemical potential $\mu$ in $E_{\bm p}=\sqrt{(\varepsilon_{\bm p}-\mu)^2+\Delta^2}$ can be safely taken to be equal to the Fermi energy $\varepsilon_{\rm F}=k_{\rm F}^2/(2m)$ in the weak-coupling BCS theory, this simplification is no longer valid for the BCS-BEC crossover region, because it remarkably deviates from $\varepsilon_{\rm F}$ there\cite{NSR,SadeMelo}. Although the Yosida function in Eq. (\ref{eq3-1}) is, strictly speaking, for a mean-field BCS superfluid, this strong-coupling correction can be effectively incorporated into $\chi_{\rm Yosida}$, by replacing $\mu$ with 
\begin{equation}
{\tilde \mu}\equiv {{\tilde k}_{\rm F}^2 \over 2m},
\label{eq4-0}
\end{equation}
where the effective Fermi momentum ${\tilde k}_{\rm F}$ obeys\cite{Hanai2}
\begin{equation}
{{\tilde k}_{\rm F}^2 \over 2m}-\mu+{\rm Re}\Sigma_{11}({\tilde {\bm k}}_{\rm F},i\omega_n\rightarrow 0+i\delta)=0,
\label{eq4-1}
\end{equation}
with $\delta$ being an infinitesimally small positive number. In obtaining Eq. (\ref{eq4-1}), since the Fermi surface is absent in the superfluid phase, we have assumed a model normal Fermi gas which is described by the single-particle Green's function ${\tilde G}({\bm p},i\omega_n)$ with the same self-energy as the (1,1) component of the ETMA self-energy in Eq. (\ref{eq2-4}) as
\begin{equation}
{\tilde G}({\bm p},i\omega_n)=
{1 \over i\omega_n-(\varepsilon_{\bm p}-\mu)-\Sigma_{11}({\bm p},i\omega)}.
\label{eq4-2}
\end{equation}
For this model normal Fermi gas, we have first evaluated the single-particle dispersion from the pole of the analytic continued Green's function ${\tilde G}({\bm p},i\omega_n\to\omega+i\delta)$, and then obtained Eq. (\ref{eq4-1}) from the condition that the quasiparticle energy $\omega$ equals zero at the effective Fermi momentum ${\tilde k}_{\rm F}$.
\par
\begin{figure}[t]
\begin{center}
\includegraphics[width=6cm]{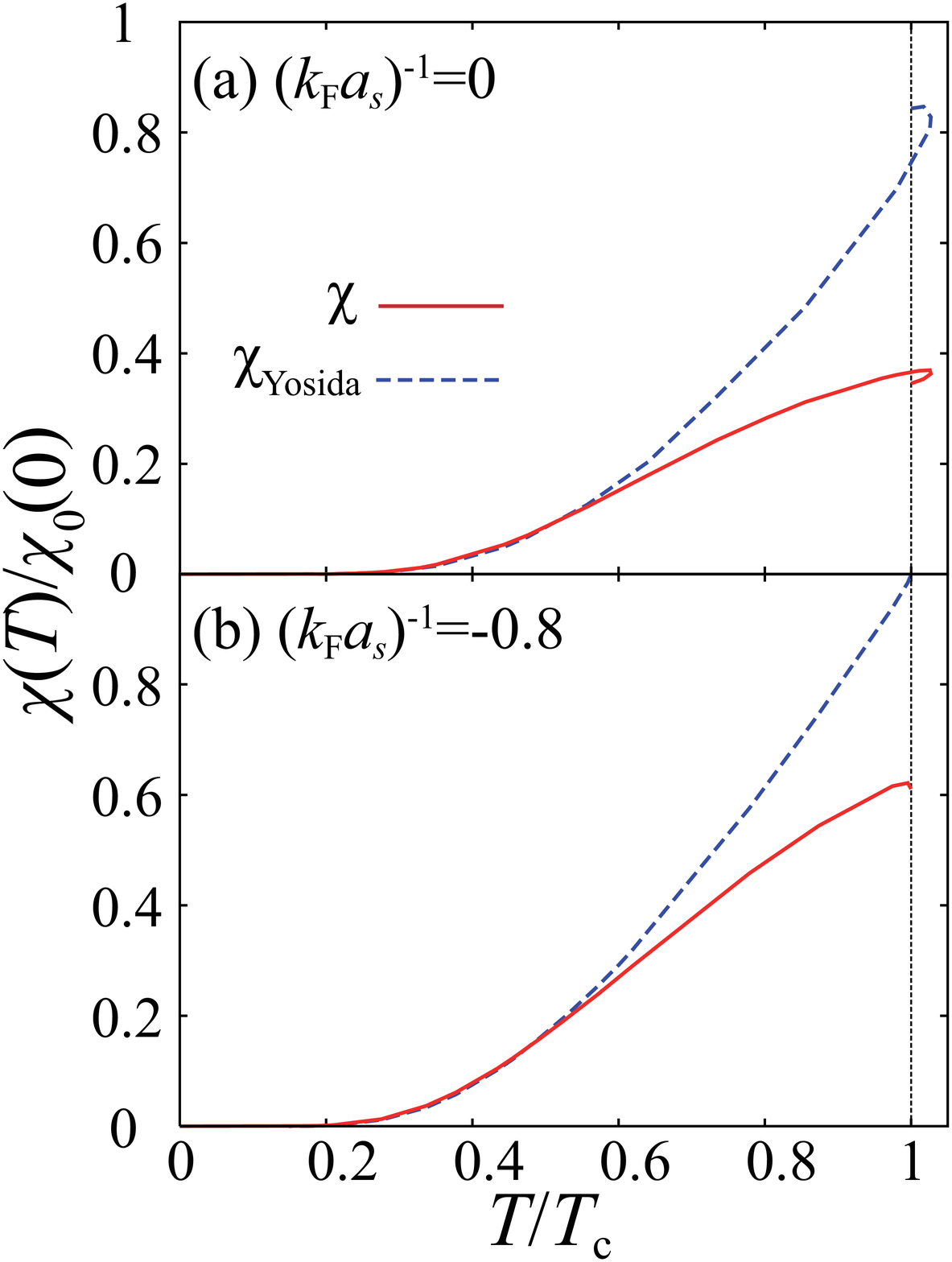}
\end{center}
\caption{(color online). Comparison of ETMA spin susceptibility $\chi$ with the Yosida function $\chi_{\rm Yosida}$\cite{Yosida}. In evaluating $\chi_{\rm Yosida}$, we use ${\tilde \mu}$ in Eq. (\ref{eq4-0}) and the ETMA superfluid order parameter $\Delta$ shown in Fig. \ref{fig2}(a).}
\label{fig6}
\end{figure}
\par
Using this replacement, and substituting the ETMA superfluid order parameter $\Delta$ into Eq. (\ref{eq3-1}), one finds in Fig. \ref{fig6} that the (modified) Yosida function $\chi_{\rm Yosida}$ well describes the low temperature behavior of $\chi$. In this temperature region, $\chi_{\rm Yosida}$ has the thermal activation type temperature dependence as,
\begin{equation}
\chi_{\rm Yosida}\sim e^{-\Delta/T},
\label{eq4-3}
\end{equation}
reflecting that spin excitations at $T\ll T_{\rm c}$ are accompanied by dissociation of Cooper pairs with a finite binding energy. Since this excitation threshold is directly related to the energy gap in the BCS superfluid density of states $\rho_{\rm BCS}(\omega)$, the agreement of ETMA $\chi$ with the (modified) mean-field result ($\chi_{\rm Yosida}$) makes us expect that the superfluid density of states $\rho(\omega)$ has a clear gap structure as in the mean-field BCS state in this low temperature region.
\par
\begin{figure}[t]
\begin{center}
\includegraphics[width=8cm]{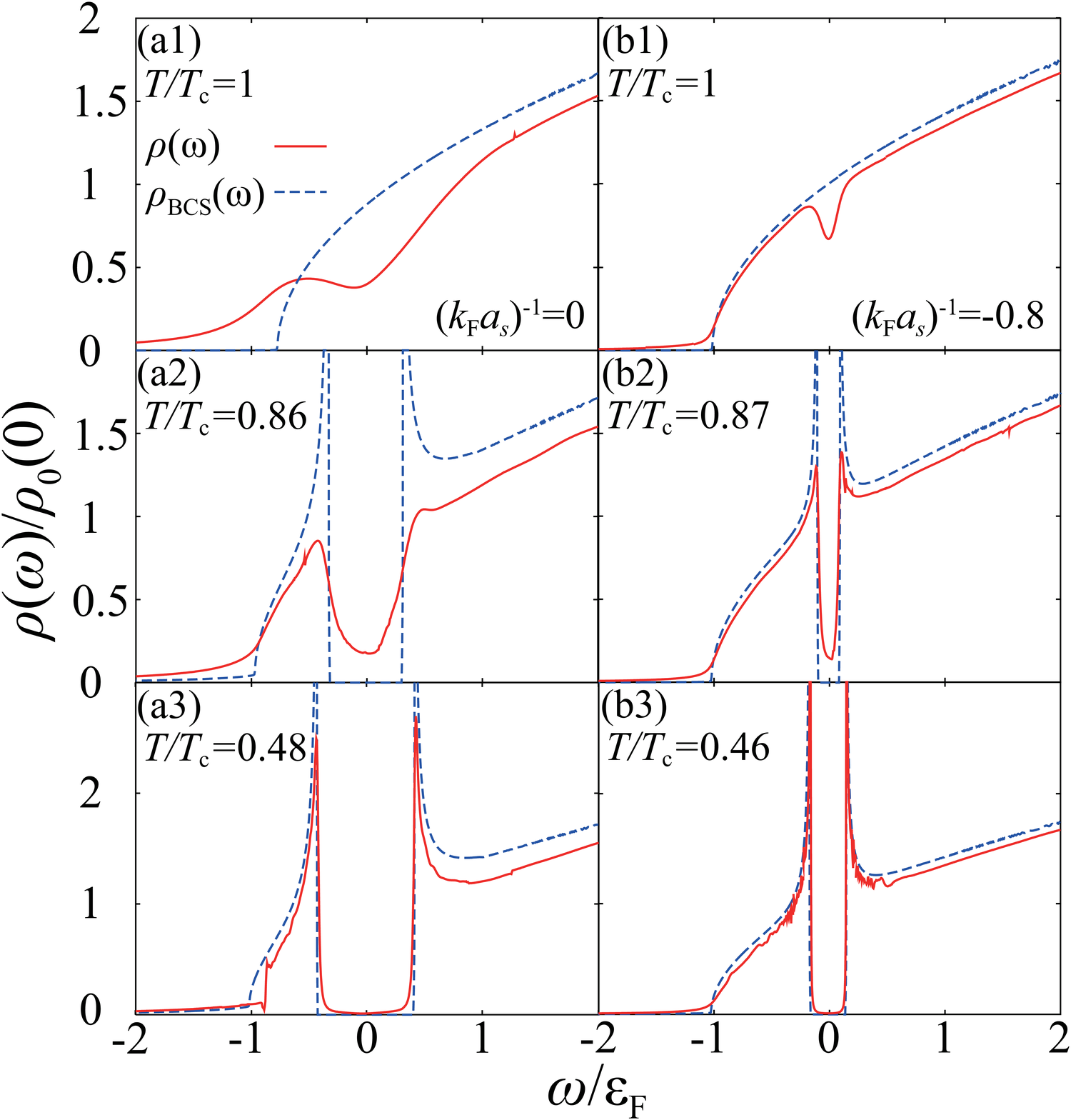}
\end{center}
\caption{(color online). Calculated superfluid density of states $\rho(\omega)$ in ETMA. (a1)-(a3) $(k_{\rm F}a_{s})^{-1}=0$ (unitarity limit). (b1)-(b3) $(k_{\rm F}a_{s})^{-1}=-0.8$. We also show the superfluid density of states $\rho_{\rm BCS}(\omega)$ in the mean-field BCS theory where the Fermi chemical potential $\mu$ is replaced by ${\tilde \mu}$ in Eq. (\ref{eq4-0}) and the ETMA superfluid order parameter $\Delta$ shown in Fig. \ref{fig2}(a) is used. $\rho_{0}(0)$ is the single-particle density of states at the Fermi level in a free Fermi gas.}
\label{fig7}
\end{figure}
\par
Figure \ref{fig7} shows $\rho(\omega)$ in a superfluid unitary Fermi gas, given by
\begin{equation}
\rho(\omega)=-{1 \over \pi}\sum_{\bm{p}}{\rm Im} G_{11}(\bm{p},i\omega_{n}\rightarrow \omega+i\delta),
\label{eq4-4}
\end{equation}
where $G_{11}(\bm{p},i\omega_{n}\rightarrow \omega+i\delta)$ is the (1,1) component of the analytic continued ETMA Green's function in Eq. (\ref{eq2-3}). As expected, we find that $\rho(\omega)$ has a almost fully gapped structure when $T/T_{\rm c}\lesssim 0.5$, where one obtains $\chi\simeq\chi_{\rm Yosida}$ (see Fig. \ref{fig6}). In the case of Fig. \ref{fig7}(a3) ($T/T_{\rm c}=0.48$), $\rho(\omega)$ is very close to the BCS result $\rho_{\rm BCS}(\omega)$, given by
\begin{eqnarray}
\rho_{\rm BCS}(\omega)
&=&
-{1 \over \pi}\sum_{\bm{p}}{\rm Im} G_0^{11}(\bm{p},i\omega_{n}\rightarrow \omega+i\delta)
\nonumber
\\
&=&\frac{m\sqrt{2m}}{4\pi^{2}}
\left[\theta(\omega-\Delta)-\theta(-\omega-\Delta)\right] \cr
&\times&
\left[\sqrt{{\tilde \mu} +\sqrt{\omega^{2}-\Delta^{2}}}\left[\frac{\omega}{\sqrt{\omega^{2}-\Delta^{2}}}-1\right]\right. \cr
&+&\left.\sqrt{{\tilde \mu} -\sqrt{\omega^{2}-\Delta^{2}}}\left[\frac{\omega} {\sqrt{\omega^{2}-\Delta^{2}}}+1\right]\theta({\tilde \mu}^{2}+\Delta^{2}-\omega^{2})\right].
\end{eqnarray}
Here, $G_0^{11}(\bm{p},i\omega_{n}\rightarrow \omega+i\delta)$ is the (1,1) component of the analytic continued mean-field Green's function in Eq. (\ref{eq2-6b}), where the effective chemical potential ${\tilde \mu}$ in Eq. (\ref{eq4-0}), as well as the ETMA $\Delta$, are used. This confirms that excitation properties far below $T_{\rm c}$ are still close to a BCS superfluid in the BCS side, as well as at the unitarity (apart from strong coupling corrections to the Fermi surface size ${\tilde k}_{\rm F}$, as well as $\Delta$).
\par
With increasing the temperature, the superfluid gap in $\rho(\omega)$ is gradually filled up, to eventually become the pseudogap at $T_{\rm c}$, as shown in Fig. \ref{fig7}\cite{Watanabe1,Watanabe2}. From the comparison of Fig. \ref{fig6} with Fig. \ref{fig7}, the deviation of $\chi$ from the Yosida function $\chi_{\rm Yosida}$ is found to be correlated to this pseudogap phenomenon in the superfluid phase. 
\par
When one simply uses the well-known knowledge for a free Fermi gas that the spin susceptibility at $T\ll T_{\rm F}$ is proportional to the density of states at the Fermi level, one might expect that the partial filling of the energy gap seen in Figs. \ref{fig7}(a2) and (b2) should enhance $\chi$ compared to the mean-field BCS case, because the latter is always accompanied by a finite energy gap below $T_{\rm c}$. However, Fig. \ref{fig6} shows the opposite result as $\chi<\chi_{\rm Yosida}$ at high temperatures ($T/T_{\rm c}\gesim 0.5$). 
\par
To understand the background physics of this phenomenon, we recall that the mean-field BCS theory treats Fermi quasi-particles (bogolons) as non-interacting particles with the Bogoliubov single-particle dispersion $E_{\bm p}$. However, the attractive interaction $-U$ should actually work between them, which would cause pairing fluctuations even below $T_{\rm c}$, where the pair-breaking and formation of spin-singlet pairs of bogolons frequently and repeatedly occur. This immediately explains the reason why the ETMA $\chi(T/T_{\rm c}\gesim 0.5)$ is smaller than $\chi_{\rm Yosida}$ in Fig. \ref{fig6}, in spite of $\rho(\omega=0)>\rho_{\rm BCS}(\omega=0)=0$. We emphasize that this mechanism is essentially the same as the preformed pair scenario for the spin-gap phenomenon above $T_{\rm c}$. In this sense, this smaller $\chi(T/T_{\rm c}\gesim 0.5)$ than $\chi_{\rm Yosida}$ may be regarded as the ``superfluid spin-gap phenomenon." Indeed, at $T_{\rm c}$, while all the Bose-condensed Cooper pairs vanish, $\chi$ is still suppressed by pairing fluctuations by non-condensed fermions, which smoothly connects to the spin-gap phase dominated by preformed Cooper pairs above $T_{\rm c}$. 
\par
We note that, in addition to the magnitude of the spin susceptibility, the temperature dependence of $\chi(T/T_{\rm c}\gesim 0.5)$ is also different from $\chi_{\rm Yosida}$ in Fig. \ref{fig6}. In the latter mean-field case, the formation of Cooper pairs only occurs below $T_{\rm c}$, so that the spin degrees of freedom is rapidly suppressed as one enters the superfluid phase, giving the remarkable decrease of $\chi_{\rm Yosida}$ just below $T_{\rm c}$. In contrast, in the ETMA case, Bose condensed Cooper pairs below $T_{\rm c}$ partially comes from preformed spin-singlet Cooper pairs that have already existed above $T_{\rm c}$. Since this part does not cause the further suppression of the spin degrees of freedom below $T_{\rm c}$, the decrease of $\chi$ with decreasing the temperature becomes weaker than the mean-field case in the superfluid phase near $T_{\rm c}$, as shown in Fig. \ref{fig6}.
\par
\begin{figure}[t]
\begin{center}
\includegraphics[width=8cm]{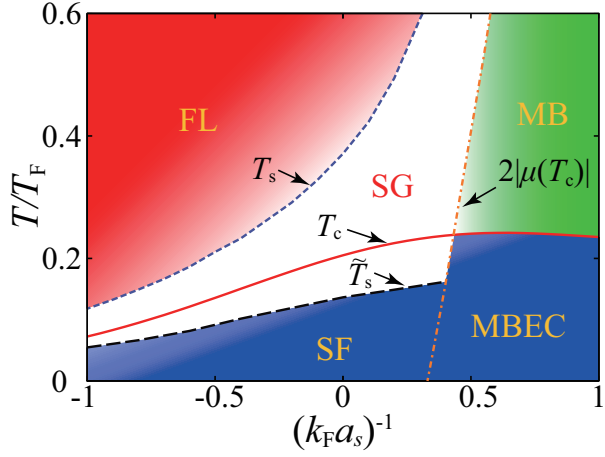}
\end{center}
\caption{(color online). Phase diagram of an ultracold Fermi gas with respect to the interaction strength $(k_{\rm F}a_s)^{-1}$ and the temperature $T$. In the superfluid phase below $T_{\rm c}$, the superfluid spin-gap temperature ${\tilde T}_{\rm s}$ is determined as the temperature at which $\chi/\chi_{\rm Yosida}=1.2$. Below ${\tilde T}_{\rm s}$ (SF), pairing fluctuations are not crucial in the sense that the spin susceptibility can be essentially described by the mean-field expression, $\chi_{\rm Yosida}$ (although we need to include strong-coupling corrections to $\mu$ and $\Delta$). Above the spin gap temperature $T_{\rm s}$ (FL), $\chi$ monotonically increases with decreasing the temperature, as in the ordinary Fermi liquid. The region between $T_{\rm s}$ and ${\tilde T}_{\rm s}$ is the spin-gap regime (SG), where pairing fluctuations crucially affect $\chi$. In this phase diagram, we also plot the line $|2\mu(T=T_{\rm c})|$ in the strong-coupling BEC regime when $\mu<0$. The region below this line may be regarded as a gas of two-body bound molecules, rather than an interacting Fermi gas. In this regime, ``NB" is the normal Bose gas regime, and ``MBEC" is the region of molecular BEC. In this phase diagram, $T_{\rm c}$ is only the phase transition temperature. The others are all characteristic temperatures, without being accompanied by any phase transition.}
\label{fig8}
\end{figure}
\par
We also note that fluctuating spin-singlet pairs of bogolons also give a finite lifetime of Bogoliubov excitations, leading to a finite width of their spectral peaks. This phenomenon smears the energy gap structure in $\rho(\omega)$, giving the ``pseudogapped" superfluid density of states shown in Figs. \ref{fig7}(a2) and (b2). Because of this, as mentioned previously, the pseudogapped superfluid density of states $\rho(\omega)$ shown in these figures and the spin-gapped spin susceptibility $\chi$ ($<\chi_{\rm Yosida}$) seen in Fig. \ref{fig6} are considered as different aspects of the same many-body phenomenon associated with pairing fluctuations in a superfluid Fermi gas.
\par
Although there is no clear phase boundary between the superfluid spin-gap regime (where $\chi<\chi_{\rm Yosida}$ and $\rho(\omega=0)>0$) and the BCS-like superfluid phase where $\chi\simeq\chi_{\rm Yosida}$ and $\rho(\omega=0)\simeq 0$, it is still convenient to introduce a characteristic temperature to physically distinguish between the two regions. As an example, plotting the temperature ${\tilde T}_{\rm s}$ at which $\chi_{\rm Yosida}/\chi=1.2$, we obtain the ``phase boundary" between the spin-gap regime (SG) and the superfluid regime (SF), as shown in Fig. \ref{fig8}. Together with our previous work above $T_{\rm c}$\cite{Tajima}, the spin-gap regime is conveniently identified as the region between the superfluid spin-gap temperature ${\tilde T}_{\rm s}$ ($<T_{\rm c}$) and the spin-gap temperature $T_{\rm s}$ ($>T_{\rm c}$) in the phase diagram in Fig. \ref{fig8}. Here, $T_{\rm s}$ is determined as the temperature at which $\chi$ takes a maximum value\cite{Tajima}. Above $T_{\rm s}$ (``FL" in Fig. \ref{fig8}), $\chi$ increases with decreasing the temperature as in the case of the ordinary normal Fermi liquid. 
\par
In Fig. \ref{fig8}, we also plot $|2\mu(T_{\rm c})|$ in the strong-coupling BEC regime where $\mu(T_{\rm c})<0$\cite{Tsuchiya1,Watanabe1,Watanabe2}. Since $|2\mu|$ equals the binding energy $E_{\rm bind}=1/(ma_s^2)$ of a two-body bound molecule in the extreme BEC limit, this quantity has the meaning of the characteristic temperature below which two-body bound molecules start to be formed, overwhelming thermal dissociation\cite{Tsuchiya1,Watanabe1}. Thus, the system below $|2\mu(T_{\rm c})|$ would be rather close to a molecular Bose gas, rather than an interacting Fermi gas. In this molecular regime, the region above $T_{\rm c}$ may be viewed as a normal Bose gas (NB), and the region below $T_{\rm c}$ is in the molecular BEC phase (MBEC)\cite{Tsuchiya1,Watanabe1}.
\par
In the phase diagram in Fig. \ref{fig8}, $T_{\rm c}$ is only the phase transition temperature. ${\tilde T}_{\rm s}$, $T_{\rm s}$, as well as $|2\mu(T_{\rm c})|$, are all characteristic temperatures, without being accompanied by any phase transition. As a result, the boundary of the spin-gap regime (SG) in Fig. \ref{fig8} somehow involves ambiguity. However, apart from this, the present spin-gap regime in the phase diagram of an ultracold Fermi gas in terms of the interaction strength and the temperature is consistent with the previous pseudogap regime determined from the pseudogapped density of states\cite{Watanabe1,Watanabe2}. As mentioned previously, while $\rho(\omega)$ is still difficult to observe in cold Fermi gas physics, the spin susceptibility $\chi$ is experimentally accessible both above and below $T_{\rm c}$\cite{Sanner}. Thus, we expect that $\chi$ is a useful quantity to experimentally explore the region where the preformed pair scenario works in the BCS-BEC crossover regime of an ultracold Fermi gas. 
\par
\par
\section{Summary}
\par
To summarize, we have discussed the spin susceptibility $\chi$ in a strongly interacting superfluid Fermi gas. Including pairing fluctuations within the framework of an extended $T$-matrix approximation, we have examined how strong-coupling corrections to this magnetic quantity are affected by the superfluid order in the BCS-BEC crossover region.
\par
In the BCS side ($(k_{\rm F}a_s)^{-1}\lesssim 0$), we showed that $\chi$ decreases with decreasing the temperature below the spin-gap temperature $T_{\rm s}$, and the suppression becomes more remarkable below $T_{\rm c}$. This is because, while the spin-gap phenomenon above $T_{\rm c}$ is dominated by fluctuating preformed spin-singlet Cooper pairs, they start to become stable below $T_{\rm c}$, which accelerates the suppression of spin excitations.
\par
In the strong-coupling BEC regime, on the other hand, tightly bound spin-singlet molecules have already been formed above $T_{\rm c}$, and $\chi$ is dominated by their thermal dissociation both above and below $T_{\rm c}$. Because of this, while $\chi$ in this regime also decreases with decreasing the temperatures as in the BCS side near $T_{\rm s}$, the detailed temperature dependence is not sensitive to the superfluid phase transition.
\par
In the BCS side of the BCS-BEC crossover region, we found that $\chi$ is well described by the Yosida function $\chi_{\rm Yosida}$ far below $T_{\rm c}$, when strong-coupling corrections to the Fermi surface size and the superfluid order parameter are appropriately taken into account. In this low temperature region, the superfluid density of states $\rho(\omega)$ has a clear energy gap as in the mean-field BCS state. Thus, apart from the above-mentioned strong-coupling corrections, superfluid properties far below $T_{\rm c}$ are essentially the same as the ordinary BCS superfluid, even in a unitary Fermi gas. 
\par
However, $\chi$ gradually becomes smaller than the mean-field result $\chi_{\rm Yosida}$ with increasing the temperature. This is because Fermi quasi-particles excited thermally (bogolons) attractively interact with each other, to cause pairing fluctuations, where formation and pair-breaking of spin-singlet pairs frequently and repeatedly occur. This strong-coupling effect is completely ignored in the mean-field theory (where bogolons are treated as non-interacting fermions). As a result, the preformed pair scenario works in the superfluid phase near $T_{\rm c}$, giving the smaller $\chi$ than the BCS result, $\chi_{\rm Yosida}$. 
\par
As a useful characteristic temperature to physically distinguish between the low temperature superfluid region where $\chi\simeq\chi_{\rm Yosida}$ and the high temperature superfluid region where pairing fluctuations crucially affect $\chi$, we conveniently introduced a superfluid spin-gap temperature ${\tilde T}_{\rm s}$. We then determined the spin-gap regime, where pairing fluctuations crucially affects $\chi$, as the region surrounded by ${\tilde T}_{\rm s}$, the spin-gap temperature $T_{\rm s}$, and $|2\mu(T_{\rm c})|$ (which physically gives the boundary between the Fermi gas regime and the molecular Bose gas regime), in the phase diagram of an ultracold Fermi gas with respect to the interaction strength and the temperature. This spin-gap regime is consistent with the previous pseudogap regime that was determined from the pseudogapped density of states\cite{Watanabe1}.
\par
In this paper, we have assumed a uniform superfluid Fermi gas, although the real system is always confined in a trap potential. Since the recent experiment on the spin susceptibility in a $^6$Li Fermi gas\cite{Sanner} is based on a local measurement of spin fluctuations, it is an interesting future problem to clarify how the spatial inhomogeneity affects the spin-gap phenomenon discussed in this paper. Since the direct observation of the single-particle density of states seems difficult in the current stage of cold Fermi gas physics, our results would be useful for the study of pseudogap phenomenon using the spin susceptibility, which is experimentally accessible both above and below $T_{\rm c}$.
\par
\acknowledgements
\par
We thank T. Kashimura, R. Watanabe, D. Inotani and P. van Wyk for useful discussions. This work was supported by KiPAS project in Keio University. HT and RH were supported by a Grant-in-Aid for JSPS fellows. YO was also supported by Grant-in-Aid for Scientific research from MEXT and JSPS in Japan (No.25400418, No.15H00840).
\par


\begin{thebibliography}{9}
\bibitem{Eagles} D. M. Eagles, Phys. Rev. \textbf{186}, 456 (1969).
\bibitem{Leggett} A. J. Leggett, in {\it Modern Trends in the Theory of Condensed Matter}, edited by A. Pekalski and J. Przystawa (Springer Verlag, Berlin, 1980), p. 14.
\bibitem{NSR} P. Nozi{\`e}res and S. Schmitt-Rink, J. Low Temp. Phys. \textbf{59}, 195 (1985).
\bibitem{SadeMelo} C. A. R. Sa de Melo, M. Randeria, and J. R. Engelbrecht, Phys. Rev. Lett. \textbf{71}, 3202 (1993).
\bibitem{Perali2} A. Perali, P. Pieri, G. C. Strinati, and C. Castellani, Phys. Rev. B \textbf{66}, 024510 (2002).
\bibitem{Ohashi} Y. Ohashi  and A. Griffin, Phys. Rev. Lett. \textbf{89}, 130402 (2002).
\bibitem{Ohashi2003} Y. Ohashi  and A. Griffin, Phys. Rev. A \textbf{67}, 063612 (2003).
\bibitem{Jin} C. A. Regal, M. Greiner, and D. S. Jin, Phys. Rev. Lett. {\bf 92}, 040403 (2004).
\bibitem{Zwierlein} M. W. Zwierlein, C. A. Stan, C. H. Schunck, S. M. F. Raupach, A. J. Kerman, and W. Ketterle, Phys. Rev. Lett. {\bf 92}, 120403 (2004).
\bibitem{Kinast} J. Kinast, S. L. Hemmer, M. E. Gehm, A. Turlapov, and J. E. Thomas, Phys. Rev. Lett. {\bf 92}, 150402 (2004). 
\bibitem{Jochim} M. Bartenstein, A. Altmeyer, S. Riedl, S. Jochim, C. Chin, J. H. Denschlag, and R. Grimm, Phys. Rev. Lett. {\bf 92}, 203201 (2004).
\bibitem{Bloch} I. Bloch, J. Dalibard, and W. Zwerger, Rev. Mod. Phys. \textbf{80}, 885 (2008).
\bibitem{Gurarie} V. Gurarie, and L. Radihovsky, Ann. Phys. \textbf{322}, 2 (2007).
\bibitem{Haussmann2} R. Haussmann, W. Rantner, S. Cerrito, and W. Zwerger, Phys. Rev. A \textbf{75}, 023610 (2007).
\bibitem{Giorgini} S. Giorgini, L. P. Pitaevskii, and S. Stringari, Rev. Mod. Phys. \textbf{80}, 1215 (2008).
\bibitem{Chin} C. Chin, R. Grimm, P. Julienne, and E. Tiesinga, Rev. Mod. Phys. \textbf{82}, 1225 (2010).
\bibitem{Heiselberg} H. Heiselberg, Phys. Rev. A \textbf{63}, 043606 (2001).
\bibitem{Carlson} J. Carlson, S.-Y. Chang, V. R. Pandharipande, and K. E. Schmidt, Phys. Rev. Lett. \textbf{91}, 050401 (2003).
\bibitem{Ho} T.-L. Ho, Phys. Rev. Lett. \textbf{92}, 090402 (2004). 
\bibitem{Hu2} H. Hu, P. D. Drummond, and X.-J. Liu, Nat. Phys. \textbf{3}, 469 (2007).
\bibitem{Nishida} Y. Nishida, Phys. Rev. A \textbf{75}, 063618 (2007).

\bibitem{Luo} L. Luo, B. Clancy, J. Joseph, J. Kinast, and J. E. Thomas, Phys. Rev. Lett. \textbf{98}, 080402 (2007).
\bibitem{Horikoshi} M. Horikoshi, S. Nakajima, M. Ueda, and T. Mukaiyama, Science \textbf{327}, 442 (2010).
\bibitem{Ku} M. J. H. Ku, A. T. Sommer, L. W. Cheuk, and M. W. Zwierlein, Science \textbf{335}, 563 (2012).
\bibitem{Pantel} P.-A. Pantel, D. Davesne, and M. Urban, Phys. Rev. A \textbf{90}, 053629 (2014).
\bibitem{Palestini} F. Palestini, P. Pieri, and G. C. Strinati, Phys. Rev. Lett. \textbf{108}, 080401 (2012). 
\bibitem{Mink} M. P. Mink, V. P. J. Jacobs, H. T. C. Stoof, R. A. Duine, M. Polini, and G. Vignale, Phys. Rev. A \textbf{86}, 063631 (2012).
\bibitem{Stewart} J. T. Stewart, J. P. Gaebler, and D. S. Jin, Nature {\bf 454}, 744 (2008).
\bibitem{Gaebler} J. P. Gaebler, J. T. Stewart, T. E. Drake, D. S. Jin, A. Perali, P. Pieri, and G. C. Strinati, Nat. Phys., \textbf{6}, 569 (2010).
\bibitem{Perali} A. Perali, F. Palestini, P. Pieri, G. C. Strinati, J. T. Stewart, J. P. Gaebler, T. E. Drake, and D. S. Jin, Phys. Rev. Lett. \textbf{106}, 060402 (2011).
\bibitem{Kohl} M. Feld, B. Fr\"{o}hlich, E. Vogt, M. Koschorreck, and M. K\"{o}hl, Nature \textbf{480}, 75 (2011).
\bibitem{Sagi} Y. Sagi, T. E. Drake, R. Paudel, R. Chapurin, and D. S. Jin, Phys. Rev. Lett. {\bf 114}, 075301 (2015).
\bibitem{Tsuchiya1} S. Tsuchiya, R. Watanabe, and Y. Ohashi, Phys. Rev. A \textbf{80}, 033613 (2009).
\bibitem{Tsuchiya2} S. Tsuchiya, R. Watanabe, and Y. Ohashi, Phys. Rev. A \textbf{82}, 033629 (2010).
\bibitem{Tsuchiya3} S. Tsuchiya, R. Watanabe, and Y. Ohashi, Phys. Rev. A \textbf{84}, 043647 (2011).
\bibitem{Watanabe1} R. Watanabe, S. Tsuchiya, and Y. Ohashi, Phys. Rev. A \textbf{82}, 043630 (2010).
\bibitem{Watanabe2} R. Watanabe, S. Tsuchiya, and Y. Ohashi, Phys. Rev. A \textbf{86}, 063603 (2012).
\bibitem{Watanabe3} R. Watanabe, S. Tsuchiya, and Y. Ohashi, Phys. Rev. A \textbf{88}, 013637 (2013).
\bibitem{Mueller} E. J. Mueller, Phys. Rev. A \textbf{83}, 053623 (2011).
\bibitem{Chen} Q. J. Chen and K. Levin, Phys. Rev. Lett. \textbf{102}, 190402 (2009).
\bibitem{Hu} H. Hu, X.-J. Liu, P. D. Drummond, and H. Dong, Phys. Rev. Lett. \textbf{104}, 240407 (2010). 
\bibitem{Magierski} P. Magierski, G. Wlazlowski, and A. Bulgac, Phys. Rev. Lett. \textbf{107}, 145304 (2011).
\bibitem{Su} S-Q. Su, D. E. Sheehy, J. Moreno, and M. Jarrell, Phys. Rev. A \textbf{81}, 051604(R) (2010).
\bibitem{Bulgac} G. Wlaz\l owski, P. Magierski, J. E. Drut, A. Bulgac, and K. J. Roche, Phys. Rev. Lett. {\bf 110}, 090401 (2013).
\bibitem{Boettcher} I. Boettcher, J. Braun, T. K. Herbst, J. M. Pawlowski, D. Roscher, and C. Wetterich, Phys. Rev. A \textbf{91}, 013610 (2015).
\bibitem{Renner} Ch. Renner, B. Revaz, J.-Y. Genoud, K. Kadowaki, and \O. Fischer, Phys. Rev. Lett. \textbf{80}, 149 (1998).
\bibitem{Krasnov} V. M. Krasnov, A. Yurgens, D. Winkler, P. Delsing, and T. Claeson, Phys. Rev. Lett. \textbf{84}, 5860 (2000). 
\bibitem{Fauque} B. Fauqu\'{e}, Y. Sidis, V. Hinkov, S. Pailh\`{e}s, C. T. Lin, X. Chaud, and P. Bourges, Phys. Rev. Lett. \textbf{96}, 197001 (2006).
\bibitem{Fischer} \O. Fischer, M. Kugler, I. Maggio-Aprile, and C. Berthod, Rev. Mod. Phys. \textbf{79}, 353 (2007).
\bibitem{Ma} J.-H. Ma, Z.-H. Pan, F. C. Niestemski, M. Neupane, Y.-M. Xu, P. Richard, K. Nakayama, T. Sato, T. Takahashi, H.-Q. Luo, L. Fang, H.-H. Wen, Z. Wang, H. Ding, and V. Madhavan, Phys. Rev. Lett. \textbf{101}, 207002 (2008).
\bibitem{Hashimoto} M. Hashimoto, R.-H. He, K. Tanaka, J.-P. Testaud, W. Meevasana, R. G. Moore, D. Lu, H. Yao, Y. Yoshida, H. Eisaki, T. P. Deveraux, Z. Hussain, and Z.-X. Shen, Nat. Phys. \textbf{6}, 414 (2010).
\bibitem{Hashimoto2} M. Hashimoto, E. A. Nowadnick, R.-H. He, I. M. Vishik, B. Moritz, Y. He, K. Tanaka, R. G. Moore, D. Lu, Y. Yoshida, M. Ishikado, T. Sasagawa, K. Fujita, S. Ishida, S. Uchida, H. Eisaki, Z. Hussain, T. P. Devereaux, and Z.-X. Shen, Nat. Mater. \textbf{14}, 37 (2015).
\bibitem{Pines} D. Pines, Z. Phys. B {\bf 103}, 129 (1997).
\bibitem{Kamp} A. Kampf and J. R. Schrieffer, Phys. Rev. B {\bf 41}, 6399 (1990).
\bibitem{Chakravarty} S. Chakravarty, R. B. Laughlin, D. K. Morr, and C. Nayak, Phys. Rev. B {\bf 63}, 094503 (2001).
\bibitem{Yanase} Y. Yanase and K. Yamada, J. Phys. Soc. Jpn. {\bf 70}, 1659 (2001).
\bibitem{Shen} A. Damascelli, Z. Hussain, and Z.-X. Shen, Rev. Mod. Phys. {\bf 75}, 473 (2003).
\bibitem{Kivelson} S. A. Kivelson, I. P. Bindloss, E. Fradkin, V. Oganesyan, J. M. Tranquada, A. Kapitulnik, and C. Howald, Rev. Mod. Phys. \textbf{75}, 1201 (2003).
\bibitem{Lee2} P. Lee, N. Nagaosa, and X. Wen, Rev. Mod. Phys. {\bf 78}, 17 (2006).
\bibitem{Nascimbene2} S. Nascimb\`{e}ne, N. Navon, K. J. Jiang, F. Chevy and C. Salomon, Nature \textbf{463}, 1057 (2010).
\bibitem{Nascimbene} S. Nascimb\`ene, N. Navon, S. Pilati, F. Chevy, S. Giorgini, A. Georges, and C. Salomon, Phys. Rev. Lett. \textbf{106}, 215303 (2011). 
\bibitem{Navon} N. Navon, S. Nascimb\`ene, F. Chevy, and C. Salomon, Science \textbf{328}, 729 (2010). 
\bibitem{Tajima} H. Tajima, T. Kashimura, R. Hanai, R. Watanabe, and Y. Ohashi, Phys. Rev. A \textbf{89}, 033617 (2014).
\bibitem{Kashimura1} T. Kashimura, R. Watanabe, and Y. Ohashi, Phys. Rev. A \textbf{86}, 043622 (2012).
\bibitem{Kashimura2} T. Kashimura, R. Watanabe, and Y. Ohashi, Phys. Rev. A \textbf{89}, 013618 (2014).
\bibitem{Hanai} R. Hanai, T. Kashimura, R. Watanabe, D. Inotani, and Y. Ohashi, Phys. Rev. A \textbf{88}, 053621 (2013).
\bibitem{Sanner} C. Sanner, E. J. Su, A. Keshet, W. Huang, J. Gillen, R. Gommers, and W. Ketterle, Phys. Rev. Lett. \textbf{106}, 010402 (2011).
\bibitem{Sommer} A. Sommer, M. Ku, G. Roati, and M. W. Zwierlein, Nature (London) \textbf{472}, 201 (2011).
\bibitem{Esslinger} J. Meineke, J.-P. Brantut, D. Stadler,T. M\"uller, H. Moritz, and T. Esslinger, Nat. Phys. \textbf{8}, 454 (2012).
\bibitem{Lee} Y.-R. Lee, T. T. Wang, T. M. Rvachov, J.-H. Choi, W. Ketterle, M.-S. Heo, Phys. Rev. A \textbf{87}, 043629 (2013).
\bibitem{Yoshinari} Y. Yoshinari, H. Yasuoka, Y. Ueda, K. Koga, and K. Kosuge, J. Phys. Soc. Jpn. \textbf{59}, 3698 (1990). 
\bibitem{Yosida} K. Yosida, Phys. Rev. Lett. \textbf{110}, 769 (1958).
\bibitem{Schrieffer} J. R. Schrieffer, {\it Theory of Superconductivity} (Addison-Wesley, NY, 1964).
\bibitem{Ohashi2} Y. Ohashi and S. Takada, J. Phys. Soc. Jpn. {\bf 66}, 2437 (1997).
\bibitem{Pieri} P. Pieri, L. Pisani, and G. C. Strinati, Phys. Rev. B \textbf{70}, 094508 (2004).
\bibitem{Fukushima} N. Fukushima, Y. Ohashi, E. Taylor, and A. Griffin, Phys. Rev. A \textbf{75}, 033609 (2007). 
\bibitem{note1} Since $\Delta$ is chosen to be parallel to the $\tau_1$ component, we actually need to subtract the static contribution $\sum_{\bm p}\langle \Psi_{\bm p}^\dagger\tau_1\Psi_{\bm p}\rangle$ from $\rho_1({\bm q}=0,i\nu_n=0)$ in the superfluid phase\cite{Ohashi2}. 
\bibitem{Haussmann} R. Haussmann, M. Punk, and W. Zwerger, Phys. Rev. A {\bf 80}, 063612 (2009).
\bibitem{OhashiJPSJ} Y. Ohashi, J. Phys. Soc. Jpn. {\bf 74}, 2659 (2005). 
\bibitem{Ward} P. Nozi\`eres, {\it Theory of Interacting Fermi Systems} (Benjamin, New York, 1964), Chap. 6.
\bibitem{Liu} X.-J. Liu and H. Hu, Europhys. Lett. \textbf{75}, 364 (2006).
\bibitem{Parish} M. M. Parish, F. M. Marchetti, A. Lamacraft, and B. D. Simons, Nat. Phys. \textbf{3}, 124 (2007).
\bibitem{note2} In the second line in Fig. \ref{fig3}(a), the diagrams with spin vertex correction are sometimes referred to as the Maki-Thompson (MT) diagrams, that are obtained from the derivative of the dressed Green's function ${\hat G}$ in the ETMA self-energy ${\hat \Sigma}$ in Eq. (\ref{eq2-4}) with respect to the effective magnetic field $h$. In addition to these, the derivative of the particle-particle scattering matrix ${\hat \Gamma}$ in ${\hat \Sigma}$ with respect to $h$ also gives the so-called Aslamazov-Larkin (AL) diagrams. However, this contribution identically vanishes in the present case, both above and below $T_{\rm c}$.
\bibitem{Enss} T. Enss and R. Haussmann, Phys. Rev. Lett. \textbf{109}, 195303 (2012).

\bibitem{Kubo} R. Kubo, {\it Statical mechanics} (North Holland, Amsterdam, 1988), Chap. 4.
\bibitem{note3} $\chi_{\rm BF}$ in Eq. (\ref{eqBF3}) is the same as the ordinary expression for the spin susceptibility in a free Fermi gas\cite{Kubo}, because the spinless bosons do not affect the spin susceptibility in the present model Bose-Fermi mixture. 
\bibitem{Hanai2} R. Hanai and Y. Ohashi, Phys. Rev. A \textbf{90}, 043622 (2014).
\end{thebibliography}
\end{document}